\documentclass{webofc}
\usepackage[varg]{txfonts}   
\usepackage{placeins}
\usepackage[subtle]{savetrees}  
\begin{document}
\title{\vspace{-0.8cm}Correlation between heavy flavour production and multiplicity in pp and p-Pb collisions at high energy in the multi-pomeron exchange model\vspace{-0.3cm}}
%
%

\author{{Grigory} \lastname{Feofilov}\inst{1} \and
        {Vladimir} \lastname{Kovalenko}\inst{1}$^\text{,a}$\footnote[0]{$^\text{a}$\email{v.kovalenko@spbu.ru}} \and
        {Andrei} \lastname{Puchkov}\inst{1}
}

\institute{Saint Petersburg State University, Russia }

\abstract{%
 The multiplicity dependence of heavy flavour production in pp-collisions at LHC energies is studied in the framework of the multi-pomeron exchange model.
The model is introducing the string-string interaction collectivity effects in pp collisions,
which modifies multiplicity and transverse momenta, leading to the non-trivial mean $p_t$ vs. multiplicity (${\left\langle
{p}_{t}\right\rangle
}_{{N}_{\text{ch}}}-{N}_{\text{ch}}$)
correlation. The string collectivity strength parameter is fixed by experimental data on multiplicity and transverse momentum correlation in a wide energy range (from ISR to LHC). The particles discrimination is implemented according to Schwinger mechanism taking into account the strong decays of hadron resonances.
We demonstrate, that the faster-than-linear growth of the open charm production with the event 
charged particle multiplicity, observed in experimental pp high energy collisions, can be explained by the modification of the string tension
due to the increasing overlap and interaction of quark-gluon strings. The model is extended for p-A interactions and the calculations for p-Pb collisions are performed.  

}

\maketitle
\section{Introduction}
\label{intro}

Measurements of open charm production  are traditionally considered as a 
probe for hot and dense matter formed in heavy ion collisions
at high energy. Charm quarks, produced in hard collisions, can interact with QCD medium, and the final distribution of the hadrons with open and hidden charm reflects the transport properties of the QGP. 
However, the mechanism of the heavy quark production is not fully understood. Perturbation QCD results have large uncertainties leaving a room for alternative mechanisms of charm production.
In this connection proton-proton interaction is considered as a benchmark for the analysis
of proton-nucleus and heavy ion collisions. 

The recent measurements on multiplicity dependence of the heavy flavour hadron production, performed by the ALICE Collaboration~\cite{ACE}, demonstrated quite unexpected results.
It was obtained that the D meson yield $N_{\text{D}}$, plotted as a function of charged multiplicity $N_{\text{ch}}$ in relative variables, shows faster-than-linear behaviour. 
This data have been interpreted in the simple string percolation model \cite{Ferreiro:2015gea}  where it was assumed that 
the bulk multiplicity decreases due to strings interaction and fusion. Later it was checked \cite{vechk} in more realistic Monte Carlo model with string fusion \cite{MC1,MC2}, where the string configurations are simulated event-by-event.
Similar effects appear in EPOS 3 model \cite{ProceedingsWerner:2016tff} with the hydrodynamical stage of the evolution.

In this paper, we calculate the yield of D mesons as a function of multiplicity in the Extended Multi-Pomeron Exchange Model \cite{EPEM,EPEM1,EPEM2}. The parameters of the model are fixed by experimental data on total (unidentified) charged multiplicity and correlation between transverse momentum with $N_{\text{ch}}$. The particle differentiation is implemented according to modified Schwinger mechanism \cite{schwinger}.

\section{Model description}
\label{sec-1}

According to the Regge-Gribov multi-pomeron exchange approach,
several pomeron exchanges in one pp-collision can happen, and the probability $w_n$ of $n$ pomeron exchanges \cite{Kaidalov,Kaidalov1} is the following:
\begin{equation}\label{prob-w_n}
w_n=\sigma_n / \sum\limits_{n'} \sigma_{n'},
\hspace{1cm}\text{where}\hspace{1cm}
\sigma_n=\frac{\sigma_P}{n z} \left( 1-e^{-z}\sum\limits_{l=0}^{n-1} \frac{z^l}{l!}\right), \hspace{1cm} z=\frac{2C\gamma s^\Delta}{R^2+\alpha'\log(s)}
\end{equation}

We used the following numerical values of the parameters, which were tuned to describe total and elastic cross sections in high energy range \cite{smth}:
\begin{equation}\nonumber
\Delta = 0.139,\ \hspace{0.4cm} \alpha'=0.21 \text{ GeV}^{-2},\ \hspace{0.4cm} 
 \gamma=1.77 \text{ GeV}^{-2},\ \hspace{0.4cm}
  R_0^2=3.18 \text{ GeV}^{-2},\ \hspace{0.4cm}  C=1.5\ .
\end{equation}

Each pomeron corresponds to a pair of strings \cite{Kaidalov,Kaidalov1}. The probability of the production of $N_{\text{ch}}$ particles from an event with $n$ pomerons is given by Poisson distribution.
The distribution of multiplicity and the transverse momentum of the hadrons of different types appears as a linear combination of the corresponding one-particle functions.

Accoring to the modified Schwinger mechanism \cite{schwinger,EPEM} in case of exchange of $n$ pomerons the probability of the production of a primary hadron of the type
$\text{$\nu $}$ with transverse momentum 
${p}_{t}$ is proportional to the value
\begin{equation*}
\text{exp}\left(-\frac{\text{$\pi
$}\left({{{p}}_{t}}^{2}+{{{m}}_{\nu
}}^{2}\right)}{{{n}}^{\text{$\beta $}}{t}}\right),
\end{equation*}
where ${{m}}_{\text{$\nu $}}$  is a mass of the particle, $t$ is the  tension of a single string, and  $\beta$ is the parameter, responsible for the collectivity. Thus in $n^\beta t$ it takes into account the increase of string tension due to the string fusion mechanism \cite{StringFusion0}. The parameter is  fixed  as a function of the collision energy from the data on ${\left\langle
{p}_{t}\right\rangle
}_{{N}_{\text{ch}}}-{N}_{\text{ch}}$ correlations in pp and $\text{p}\bar{{\text{p}}}$ collisions.
Parameter $\beta=0$ means no collectivity, and no correlation between transverse momentum and multiplicity. At $\beta>0$ $(\beta<0)$ the positive (negative) ${\left\langle
{p}_{t}\right\rangle
}_{{N}_{\text{ch}}}-{N}_{\text{ch}}$ correlation appears.
The procedure for fixing the parameters is in detail described in the \cite{EPEM1,EPEM2}.
Particularly, parameter $t$ is taken to be a constant: t~=~$0. 566$ $\text{GeV}^2$, and for parameter $\beta$ a smooth behavior  with energy has been obtained and approximated by $\beta =\ 1.16\  {(1-\log(\sqrt{s}/(1\ \text{GeV}))-2.52)}^{0.19}$. No extra parameter tuning has been made for particle differentiation.

The final hadron spectrum is modified due to the cascade decay of resonances by the strong interaction. They are effectively taken into account by a cascade branching matrix ${M}_{\mu \nu}$, extracted
from the particle decayer built into THERMINATOR 2 Monte Carlo generator \cite{therminator2}.

Finally, the relative yields of particles corrected for the decays of hadrons have the form:
\begin{equation*}
{Y}_{\nu }{\sim}\sum _{\mu }{{M}_{\mu \nu }}{\cdot}\left(2{S}_{\mu
}+1\right){\cdot}\exp \left(-\frac{\pi \left({p}_{t}^{2}+{m}_{\mu
}^{2}\right)}{{n}^{\beta }t}\right),
\end{equation*}
where  ${S}_{\mu }$ is  spin of a particle of the type  $\mu $,  ${M}_{\mu \nu }$ is effective branching matrix.
{
The yields of particles are normalized keeping the total multiplicity ${N}_{\text{ch}}$.}

\vspace{0.3cm}

\subsection{Extension of the model for p-A collisions}

The present model can be extended to the case of proton-nucleus collisions. Taking into account the fact that from the proton side in all inelastic events we have one participating nucleon, the total distribution of the number of pomerons is determined by 
the distribution of the participating nucleons of the nucleus.
From the experimental data \cite{alicepa} we know, that in p-A and d-A collisions the charged multiplicity is approximately proportional to the number of participating nucleons, so that 
$\dfrac{dN/d\eta}{\langle N_{\text{part}}\rangle}|_{p-A} \simeq \dfrac{dN/d\eta}{2}|_{pp}$.
In the present model, for p-A collisions we assume that in an event for each participant pair one emits n pomerons with probability $w_n$. Consequently, the probability of the production of $n$ pomerons in p-A collision appears as a convolution of the distribution of the number of participating nucleons with the distribution $w_n$  obtained for pp interaction (see figure~1). The distribution of $N_{\text{part}}$ was calculated in Glauber model \cite{glauber}.

\begin{figure}[h]
\centering
\includegraphics[width=15.2cm,clip]{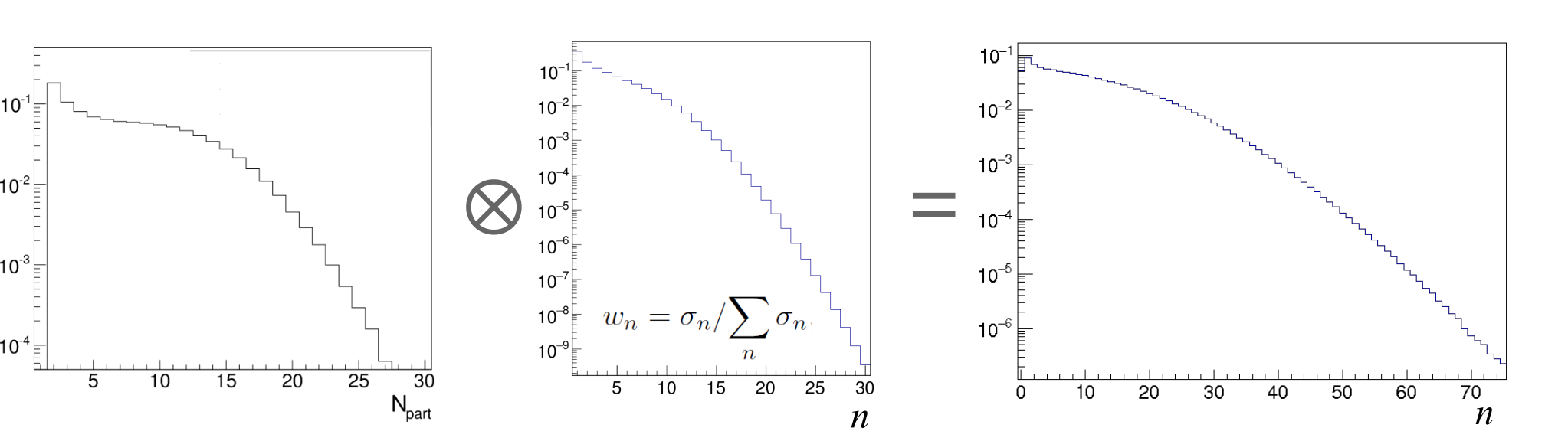}
\vspace{-0.62cm}\newline
\hspace{7cm} \textit{a} \hspace{4.5cm} \textit{b} \hspace{4.7cm} \textit{c} 
\caption{Probability of production of $n$ pomerons (\textit{c}) obtained
as a convolution of $N_\text{part}$ distribution (\textit{a}) and the
distribution of $n$ pomerons from pp interaction (\textit{b}).}\vspace{-0.54cm}
\label{fig1}       
\end{figure}

\section{Results}

In figure~2 the relative yields of strange and multi-strange hadrons as a function of the charged multiplicity are shown for pp collisions at 7 TeV \cite{strangearxiv}. The results demonstrate the fast growth of relative strangeness, similarly to what was recently observed in the ALICE experiment \cite{data}. We note that this increase of strange hadrons is obtained due to modification of string tension, similarly to the predictions of string fusion model \cite{StringFusion0}.

\begin{figure}[h]
\centering
\includegraphics[width=4.5cm,clip]{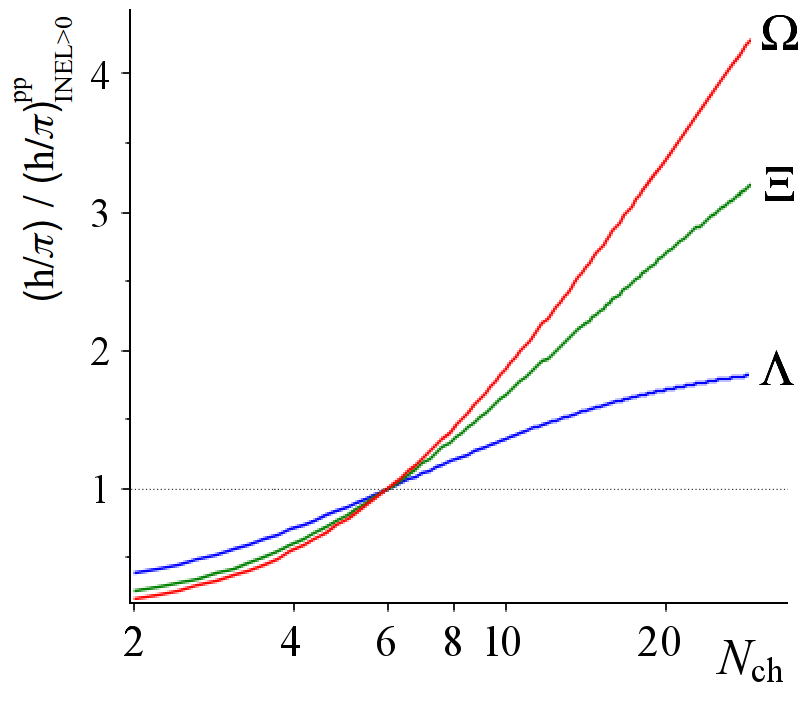}
\hspace{0.35cm}
\includegraphics[width=5.0cm,clip]{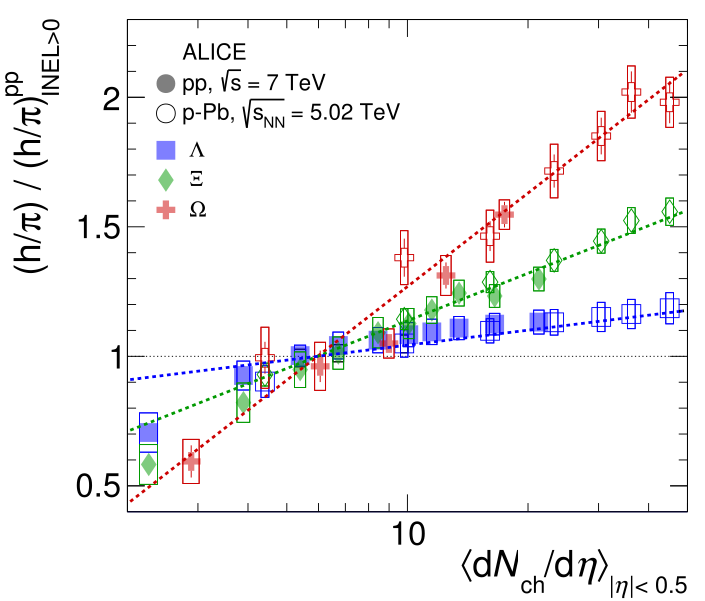}
\caption{Multiplicity dependence of the relative yields of $\Omega,\  \Xi \text{ and } \Lambda$ baryons, normalized to pion multiplicity,
in pp-collisions at $\sqrt{s}$~=7~TeV. Left plot shows the model
calculation \cite{strangearxiv}, right -- experimental data \cite{data}. Linear growth would correspond to unity, which is marked by dotted line.}
\label{fig2}       
\end{figure}

\begin{figure}[h]
\centering
\sidecaption\hspace{0.25cm}
\includegraphics[width=5.5cm,clip]{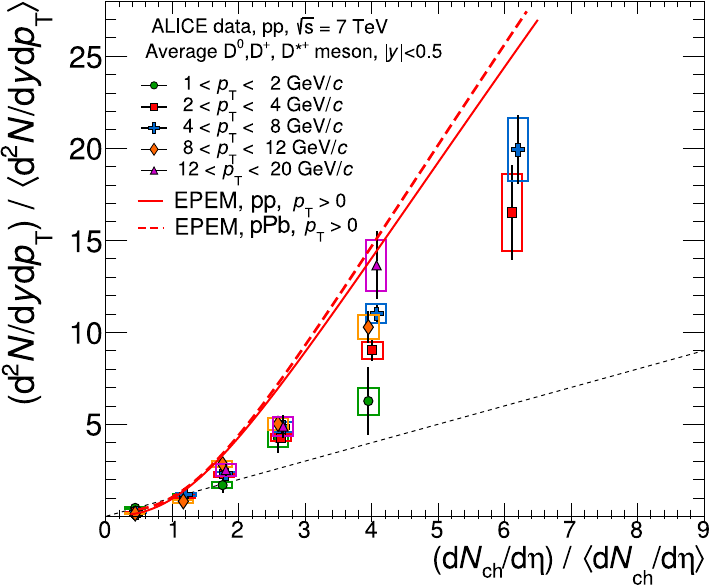}
\caption{The relative open charm yield in pp collisions at $\sqrt{s}$=7~TeV as a function of normalized charged multiplicity, calculated in extended multi-pomeron exchange model (EPEM), compared with the experimental data \cite{ACE}. The dotted line shows linear dependence of the yield (proportional to the total charged multiplicity). The solid line shows the model results for pp collisions, dashed -- for p-Pb.
}
\label{fig3}       
\end{figure}

Figure~3 shows the multiplicity dependence of open charm production in pp and p-Pb collisions at $\sqrt{s}=$~7~TeV. The results show, that the present model reproduces the faster-than-linear growth of D meson yield with multiplicity of pp collisions. The fact that the enhancement in the multi-pomeron model is even stronger than in the experiment \cite{ACE}, may be because the model collective parameter $\beta$ effectively takes into account all the collective processes leading to a correlation between the multiplicity and the transverse momentum, although the other contributions due to hard scattering processes, which are not related to the modification of the string tension \cite{effects}, also exist.
The results of the model calculation for p-Pb collisions show the dependence similar to pp case. This behavior is in agreement with the experimental data \cite{ewr}.

%

\section{Conclusions}
A generalization of the multi-pomeron exchange model with effective
account of interaction between strings is proposed allowing for the
production of charm particles in the Schwinger mechanism.
The model parameters are determined by experimental data on ${\left\langle
{p}_{t}\right\rangle
}_{{N}_{\text{ch}}}-{N}_{\text{ch}}$
correlation.
No additional parameters for particle differentiation are introduced.
The multiplicity dependence of the strange baryons and charmed meson yields calculated in the model is in qualitative agreement with experimental data.
The model is extended for the description of p-Pb collisions and
it predicts a similar yield of the open charm as in the case of pp interaction at the same multiplicity,
which is consistent with the experimental data. 

\section{Acknowledgements}
The research was supported by the grant of the Russian Science Foundation (project 16-12-10176).

%
%
%
%
%

%
%
%

\end{document}